\documentclass[preprint]{elsarticle} \usepackage{graphicx}
\usepackage{epsfig}

\begin{document}

\title{Carbon-based nanostructured composite films: elastic, mechanical
  and optoelectronic properties derived from computer simulations}

\author[au1]{Maria Fyta\corref{cor1}\fnref{mf}} 
\ead{mfyta@ph.tum.de}
\author[au1]{Christos Mathioudakis\fnref{cm}}
\ead{chrm@materials.uoc.gr}
\author[au1]{Ioannis N. Remediakis\fnref{inr}}
\ead{remed@materials.uoc.gr}
\author[au1]{Pantelis C. Kelires\fnref{pck}}
\ead{pantelis.kelires@cut.ac.cy}
\address[au1]{Physics Department, University of Crete, P.O. Box 2208,
710 03, Heraklion, 
Crete, Greece}

\fntext[mf]{Present address: Department of Physics, Technical
  University of Munich, Garching, Germany} \fntext[cm]{Present
  address: Department of Materials Science and Technology, University
  of Crete, Heraklion, Crete, Greece} \fntext[inr]{Present address:
  Department of Materials Science and Technology, University of Crete,
  Heraklion, Crete, Greece} \fntext[pck] {Present address: Department
  of Mechanical Engineering and Materials Science and Engineering,
  Cyprus University of Technology, Limassol, Cyprus}

\cortext[cor1]{Corresponding author}
%\cortext[cor2]{Principal corresponding author}

\begin{abstract}
In this review, we present our recent computational work on  carbon-based nanostructured composites. These materials consist of carbon
crystallites embedded in an amorphous carbon matrix and are modeled
here through classical and semi-empirical quantum-mechanical
simulations. We investigate the energetics, mechano-elastic, and
optoelectronic properties of these materials. Once the stability of
the composites is discussed, we move on to the calculation of their
elastic moduli and constants, their anisotropy and elastic
recovery. At  a next step, we focus on diamond composites, which were
found to be the most stable among the composites studied, and went beyond
the elastic regime to investigate their ideal fracture. Finally, for
these materials, the electronic density of states, dielectric
function, and optical response were calculated and linked to the
disorder in the structures. Our findings unveil the high
potential of these materials in nanotechnological applications,
especially as ultlra-hard coatings.
\end{abstract}

\maketitle

\section{Introduction}

Carbon-based nanostructured composites (n$a$-C) \cite{Subra98,Banhart99} are
believed to intermingle properties of carbon nanostructures with those
of single-phase amorphous carbon ($a$-C)\cite{Robertson_rev}. These nanocomposites are
hybrid forms of carbon in which nanocrystallites are embedded in the
$a$-C matrix. In this sense, their properties can be tailored by
controlling the type and size of the embedded structures, as well as the matrix material. Such composites, also investigated in this work, are relevant to both bulk and thin film materials. 
 
A variety of carbon nanostructures can compose n$a$-C, ranging from
diamond crystallites, to entirely three-dimensional sp$^2$ covalent
conformations with no dispersion/van der Waals (vdW) bonding,
i.e. open graphene structures with negative curvature (schwarzites)
\cite{Keeffe,Vanderbilt,Lenosky92}, to carbon nanotubes (CNT)
\cite{iijima} bound by vdW forces.  N$a$-C films have been synthesized
before by using different methods \cite{Amaratunga}.  The synthesis of
nanodiamond composites (nD/$a$-C) and its growth mechanisms have been
explored in both hydrogenated \cite{Lifshitz,zhou02} and pure \cite{Yao,kulisch08}
tetrahedral amorphous carbon (t$a$-C) matrices.

Here, we review our recent theoretical and simulational studies of nanostructured carbon films, which shed light on their mechanical and optoelectronic properties. This review is organized as follows: Section \ref{sec:meth} 
outlines the methodology followed here. Section \ref{sec:ene} discusses the
energetics and stability of all modeled nanocomposites. The elastic
and mechanical properties of diamond and nanotube composites are
investigated in section \ref{sec:elast}. The ideal fracture and
optoelectronic properties of diamond nanocomposites are discussed in
sections \ref{sec:fract}, \ref{sec:optoel}. We summarize in section
\ref{sec:concl}.

\section{Methodology\label{sec:meth}}

We have recently followed a theoretical approach to investigate n$a$-C and unravel the
fundamental principles governing the interaction of nanostructures
with the $a$-C matrix (see for example \cite{fytaPRB05,CNTinclusions}).
Two different methods were used, a Monte Carlo (MC) method within the
Tersoff empirical potential approach \cite{tersoff88}, and a tight-binding molecular
dynamics (TBMD) scheme which utilizes separately two hamiltonians, the
Mehl \& Papaconstantopoulos \cite{NRL1,NRL2} (TBNRL), and the
environment-dependent (EDTB) Wang \& Ho \cite{tang96} model. The former employs
non-orthogonal atomic orbitals, while the latter goes beyond the traditional two-center approximation and allows the tight-binding parameters to change according to the bonding
environment.  Using both the MC and TBMD, one is able to combine
higher statistical precision with higher accuracy and bridge the gap
between classical and first-principles calculations.  Both the Tersoff
potential, as well as the two hamiltonians have been used to successfully model the interatomic interactions in various crystalline and amorphous carbon forms
\cite{kelires94,Mathiou,Galli1,chrm04,fyta06}

The TBMD simulations are carried out in the canonical ($N,V,T$)
ensemble. The volume $V$ and the number of particles (atoms) $N$ are
constant. Dispersion/van der Waals (vdW) forces within the Tersoff potential
approach, will be included by adding a simple Lennard-Jones potential,
which successfully describes properties of carbon nanocrystals
\cite{LJparam1,LJparam2}. The parameters used for this potential are $\epsilon=
2.964~meV$ and $\sigma = 3.407~\AA$.

For the calculation of the optoelectronic properties the EDTB scheme
was used, which has proved to be quite efficient in similar cases
\cite{chrm04}. The electronic structure can be easily extracted and is
directly correlated to the imaginary part $\varepsilon_2$ of the complex
dielectric function. This is in turn yields the optical gaps and
absorption coefficient, which further link to the disorder in the
structures through the Urbach energy. Additional details can be found
elsewhere \cite{chrm_optoel_prb}.

\subsection{Structure generation\label{sec:gener}}

We have generated the nanocomposites studied here by melting and
subsequent quenching at constant volume various carbon
  crystalline cells, while keeping a certain number of atoms in the
central portion of the cells frozen in their ideal crystal
positions. Consequently, the cells are thoroughly relaxed with respect
to atomic positions and density.  For the nanotube composites, a core
is removed from a diamond crystal leaving a void with a shape similar
to that of the nanotube, in which the nanotube is then placed. Details
can be found in Ref. \cite{fytaPRB05,CNTinclusions}.

The total number of atoms in the structures ranges from $\sim$ 2500 to
4536, and the number of atoms in the nanocrystals ranges from $\sim$
300 to 900. The size of the computational cells ranges from about 1.5
to 4.0 nm.  For the TBMD relaxed structures we use smaller cells up
to 512 atoms. The diameter (d) of all crystalline inclusions used are
in the range d=0.6-2.8 nm. Periodic boundary conditions (PBC) are
applied to the cells. The exact details for each composite studied with the Tersoff potential are summarized in Table \ref{Tab:struc}. Diamond inclusions are almost spherical, while
CNTs have open ends and extend through the entire length of the
cube. Due to the PBC, this corresponds to a dense array of CNTs packed
in parallel, which is an idealized model of a CNT nanocomposite. However, this model captures all essential features of the CNT-matrix interaction resembles CNT bundles with $a$-C material in between the tubes.

Representative structures of a specific group of conformations are
chosen and embedded in amorphous matrices: the $sp^3$ conformations
(a) Diamond (D), and the (b) high-pressure phase BC8 structure for
carbon, the negatively-curved $sp^{2}$ conformations with a $D$
periodic minimal surface, (c) polybenzene (6.8$^2D$) \cite{Keeffe},
(d) PCCM (porous conducting carbon modification), an sp$^2$-bonded structure with
eightfold rings \cite{Winkler}, (e) {\it C$_{168}$}, a low-density,
negative-curvature schwarzite structure \cite{Vanderbilt}, (f) the
positively curved $sp^{2}$ carbon nanotubes (CNT) \cite{iijima} with
single (SWNT) and multiple (MWNT) graphitic walls. Structures (c)-(e)
do not involve vdW bonding. In Fig.\ref{fig:strucs}, representative
nanocomposites of the ones studied here are shown: a diamond of 1.70 nm diameter, a
$C_{168}$, and a (7,10) SWNT of 1.17 nm diameter, embedded in a highly
tetrahedral ($z_{am} \simeq$ 3.8, $\rho \simeq$ 3.3 g cm$^{-3}$) and low density ($z_{am} \simeq$ 2.8, $\rho \simeq$ 1.5 g cm$^{-3}$, and $z_{am} \simeq$ 3.2, $\rho \simeq$ 2.2 g cm$^{-3}$, respectively) matrices. The mean coordination of the amorphous matrix and the total density of the composites are denoted z$_{am}$ and $\rho$, respectively. Fig.\ref{fig:strucs}(b) shows clearly, that no covalent bonding between CNT atoms and the matrix exists, as the vdW forces drive a reconstruction of the surrounding medium, producing a
graphitic wall at a distance of 3.4 \AA ~from the CNT.

\section{Formation Energies\label{sec:ene}}

We begin with the investigation of the stability of the nanocomposites
studied here. The interaction of the embedded configuration with the
host is taken care of by defining the formation energy of a
nanocrystal, given by
\begin{equation}
E_{form} = E_{total} - N_{a}E_{a} - N_{c}E_{c}
\end{equation}
In this equation, $E_{total}$, $E_{c}$ are the total cohesive energies
per atom of the composite system and the crystalline phase,
respectively. $N_{c}$, $N_{a}$ are the number of atoms in the
nanocrystal, and the amorphous matrix, respectively, and $E_{a}$ is
the cohesive energy per atom of the pure amorphous phase (without the
nanocrystal) with coordination $z_{am}$ \cite{fytaPRB05}.  A negative
value of $E_{form}$ denotes stability of the nanostructure, a positive
value indicates metastability.

The formation energies, hence stabilities, of all nanocomposites in
this study are depicted in Fig.\ref{fig:eform}. Of the nanostructures
considered, only diamond and the C$_{168}$ schwarzite are stable
($E_{form} < 0$), the former at high $z_{am}$ ($\rho_{am} \geq
3\ gcm^{-3}$), the latter at low $z_{am}$ ($\rho_{am} \leq
1.2\ gcm^{-3}$).  Both crystallites are stable in a matrix with structural characteristics similar to their own, which relates to a small density gradient between the matrix and the inclusion: diamonds(schwarzites) are dense $sp^3$(low density $sp^2$) networks, thus are embedded in a dense(dilute) matrix. A small density gradient does not give rise to discontinuities and thereby energetically favors the embedding of the crystalline inclusion. Overall, diamonds are more stable than schwarzites. Their enhanced stability in highly tetrahedral networks
indicates that once nucleation centers are formed large diamond
regions can be further grown, as shown in experiments \cite{Lifshitz,zhou02}.
Through annealing of the diamond composites, an unstable structure
extensively shrinks, and only a small core remains intact, while the
stable structure retains its shape and should expand.  Metastable
diamonds are on the average larger, but more deformed, than stable
diamonds in t$a$-C.  For a better nanostructured material it is
preferable to nucleate diamond cores in a t$a$-C matrix.

All other nanostructures studied in this work are metastable
($E_{form} > 0$) through the whole coordination range. Their curves
exhibit a well defined local minimum at which $E_{form}$ is quite
small, suggesting that they can be maintained in the amorphous matrix
under moderate conditions and their synthesis is possible.
The fact that $E_{form}$ for 6.8$^2D$ and PCCM composites are only slightly lower than that for a diamond inclusion suggests the possibility of a phase transformation of $sp^2$-bonded structures to diamond under the proper conditions of pressure and temperature. This would subsequently aid the experiments towards synthesis of such theoretical carbon crystals. More details are discussed in  \cite{fytaPRB05}. CNT composites are also metastable with minima in their $E_{form} $ at intermediate coordinations $z_{am}$.

\section{Elastic and mechanical properties\label{sec:elast}}

The elastic and mechanical properties of carbon composites are
important for potential applications as ultra-hard protective or biomedical coatings. We focus on diamond nanocomposites ($n$D/$a$-C), which are the most
stable among the ones we dealt with in this work, and the nanotube
nanocomposites (CNT/$a$-C), as a representative sp$^2$ structure with
positive curvature which is of wide technological interest.

\subsection{Bulk and Young's moduli}

We begin with the calculations of the bulk modulus ($B$), which was
calculated by fitting the energy vs volume curve after a hydrostatic deformation to the
Birch-Murnacham equation \cite{Murnaghan}.
Figs. \ref{fig:nDmodul}(a), \ref{fig:swntmodul}(a) show the calculated
moduli plotted as a function of the nanocomposite density for various
diamond and nanotube diameters, respectively. The bulk modulus is
significantly enhanced with increasing density and inclusion size.
Replacement of some amorphous material by crystalline increases
noticeably the bulk modulus, with respect to $B$ of $a$-C, in both
diamond \cite{fyta06} and nanotube composites
(Fig.\ref{fig:swntmodul}(a)). $B$ for some samples is even higher than
that of the {\it amorphous diamond} network (WWW) \cite{www} and close
to that of diamond \cite{fyta06}. A similar enhancement of the
hardness with the grain size was found in ultra-nanocrystalline
diamond \cite{remed_uncd} and copper films \cite{galanis2010}.  A comparison of $B$ for SWNT and MWNT inclusions of the same diameters embedded in $a$-C of similar
densities reveals, that the number of walls that consist a MWNT is
crucial for the whole composite's rigidity (see
Fig.\ref{fig:swntmodul}(b)). By increasing the number of graphitic
walls within a CNT the composite becomes more rigid. 

A check whether the rule of mixtures \cite{mixtures} applies to diamond and CNT
composites revealed that it indeed holds for the former, but not for
the latter structures. Apparently, the CNT composites are significantly influenced by the presence of the matrix. In these structures, the microstructural details of the matrix have been significantly altered with respect to pure $a$-C. We note again, that the effect of the inclusion is to drive the surrounding matrix atoms further away from the nanotube and stabilize them at a distance of about 3.4 \AA. On the other hand, no such a significant effect was evident in $n$D/$a$-C, as there is a smooth structural transition from the inclusion to the matrix. In other words, there is a direct connection of the inclusion-matrix bonding and the validity of the rule of mixtures: in the nanotube composites the bonding between the matrix and the inclusion is of a van der Waals type and the rule does not hold, while in the diamond composites the two phases are covalently bonded and this rule still holds.

The trends for the Young's moduli ($Y$), which is easily extracted
from the elastic constants discussed next, of both composites are the
same as in the case of $B$.  For $n$D/$a$-C, $Y$ is shown in
Fig.\ref{fig:nDmodul}. A considerable enhancement of $Y$ was found,
especially in high density matrices, with values approaching that for
diamond \cite{fyta06}. In the case of CNT/$a$-C, a marked
characteristic is the high elastic anisotropy which is monitored by
the ratio 
\begin{equation}
A = Y_{\perp} / Y_{\parallel}
\end{equation}
where $Y_{\parallel}$,
$Y_{\perp}$ are the components of $Y$ in the direction of the tube
(axial) and in the transverse directions. A composite with a SWNT with
diameter of 1.65 nm and a density of 1.99 g cm$^{-3}$ (2.14 g
cm$^{-3}$) has $Y_{\parallel}$ = 478(564) GPa and $Y_{\perp}$ =
231(306) GPa, yielding $A$ = 0.48(0.54). The anisotropy increases with
increasing tube diameter, as the tube contribution overwhelms the
isotropic matrix part.

\subsection{Elastic constants}

To supplement the calculation of $B$, the remaining elastic moduli are
computed. All simulation cells of the diamond composites are cubic systems, thus only the
knowledge of the $c_{11}, c_{12}$ and $c_{44}$ is needed. This is not exactly the case for CNT composites, as these are modeled through tetragonal computational cells.
To calculate the elastic constants, we apply the appropriate deformation to the
system and compute its total energy as a function of the imposed
strain.  The curvature of this function at its minimum yields the
desired modulus. The results are plotted in Fig.\ref{fig:nDmodul}(b)
as a function of the density for diamond composites.  All elastic
constants increase with increasing density. This is also the case in
CNT/$a$-C, for which representative values are summarized in Table
\ref{Tab:modCNT}.  For an isotropic material,
\begin{equation}
\eta=\frac{c_{11}-c_{12}}{2c_{44}}=1
\end{equation}
which holds within 4\% or less,
for $n$D/$a$-C. For a typical CNT composite, $\eta=0.75$, while for
all pure $a$-C networks $\eta=0.98$. 

We have also looked at the elastic recovery during a
compression-decompression cycle from 0 to 100 GPa for diamond and
nanotube composites (with inclusions of sizes 1.7 nm and 1.2 nm,
respectively), as well as for amorphous structures
\cite{fyta_recov07}. The elastic recovery of $n$D/$a$-C and
single-phase $a$-C is almost perfect in the highly tetrahedral regime
($ta$-C), but is not perfect as the $sp^3$ fraction declines. An
almost fully $sp^2$-bonded $a$-C network regains its initial volume at
about 95 \%. A CNT composite, having a high sp$^2$ fraction does not
fully recover and shows a hysterisis in the compression curves. Our
results indicate that the compressibility from CNT/$a$-C towards $a$-C
and nD/$a$-C is decreasing, as long as the sp$^2$ fraction remains
high.

\section{Ideal strength of diamond nanocomposites\label{sec:fract}}

A study beyond the elastic regime gives an insight to the ideal fracture
of $a$-C and $n$D/$a$-C. Tensile (along diamond's easy slip \{111\}
plane) and shear (on the \{111\} plane in a $\langle$112$\rangle$
direction) loads were applied and the stress versus strain curves were
extracted.  The NRL hamiltonian at low temperature was used in this
study. This method has already been tested in the case of diamond and the
findings compared well with previous {\it ab initio} results
\cite{Dcleavage,Dinstab}.

Three $a$-C samples were considered (a typical non-hydrogenated $a$-C with
$z_{am}$=3.47, a t$a$-C sample with $z_{am}$=3.78 and a WWW sample)
and a $n$D/$a$-C with an inclusion of about 1.2 nm embedded in t$a$-C
with an sp$^2$ fraction of about 30\%. The stress vs. strain curves in the
case of the tensile load are summarized in Fig.\ref{fig:fract},
left panel. The maximum stresses are roughly 60, 40, 30, and 124
GPa for $z_{am}$=4.00, 3.78, 3.47, and diamond, respectively. Thus,
the ideal strength of $a$-C is proportional to its sp$^3$ ratio. The stress vs. strain
curves for the low-density $a$-C ($sp^{3}$ fraction $\sim$ 50 \%) sample suggest a ductile
behavior. The fracture of $a$-C, which has an ideal strength half that of t$a$-C, takes place at a higher load than t$a$-C, i.e. larger strain needs to be applied to break the $a$-C network. Similar are the results when a shear load is applied.

For the diamond composite, the easy slip plane of the embedded diamond
core was not found to play a crucial role in the total strength and no
weak directions were present.  The crystalline phase remains
unaffected by the external load, which is almost completely taken by
the surrounding amorphous matrix. The response, thus, of $n$D/$a$-C to
external load beyond the elastic regime is identical to the response
of the embedding matrix. The onset of fracture takes place at the weakly
bonded $sp^3$ sites in the amorphous matrix \cite{fyta06}.  Bonds in
the crystal are at this point actually elongated. Fracture in diamond
nanocomposites is inter-grain and their ideal strength is similar to
the pure amorphous phase. A snapshot of a $n$D/$a$-C at the onset of
fracture (i.e. at the maximum stress) is shown in Fig.\ref{fig:fract}, right panel.

\section{Optoelectronic properties\label{sec:optoel}}

Finally, focus is given on the electronic and optical properties of
diamond nanocomposites. We were able to probe at a local atomic level
the optoelectronic response of the composite and link it to the
associated disorder, which exists at the interface between the diamond
inclusions and the embedding $a$-C matrix. This can be quantitatively
probed by extracting the Urbach energies from the optical
parameters. Disorder in the nanocrystals appears in their outer shell near the interface and is manifested as bond length and angle distortions.

As representative quantities of an analysis done elsewhere
\cite{chrm_optoel_prb}, the electronic density of states (EDOS) and
the imaginary part ($\varepsilon_2$) of the dielectric function for nD/$a$-C are shown
in Fig.\ref{fig:opt}. The same approach for the study of the dielectric response for diamond-like amorphous carbon has been used before \cite{chrmTSFilms05}. The results compare qualitatively well with recent experiments \cite{franta2010}, taking into account that in the latter case the diamond-like film was hydrogenated.
A first look at the EDOS in Fig.\ref{fig:opt} reveals that the main
contribution to the formation of the gap, i.e. the absence of
electronic states, and of any states located in the gap region comes
from the amorphous matrix part, as this has a lower energy gap than
diamond due to the presence of sp$^2$ hybrids. As the sp$^2$ ratio,
which controls the gap through the separation of $\pi$-$\pi^\star$
bonding-antibonding states increases, the gap region shrinks. A
smaller contribution to the gap states originates from distorted
sp$^3$ atoms of the diamond inclusion at the interface.

Decomposition of the total $\varepsilon_2$ into contributions from the atoms
in the crystalline and amorphous regions showed, that the optical
response of the nanocomposite system is progressively varied. This variation occurs as the
sp$^2$ fraction in the amorphous part is increased from having nearly
diamond-like nature to having nearly graphite-like nature. It was also
clear that the energy gaps, i.e. absence of electronic transitions at
the edge of the spectra, are reduced as the sp$^2$ fraction
increases. Near the gap the main contribution comes from transitions
between states localized in atoms of the amorphous part.

Calculation of the Urbach energy (E$_U$) in $n$D/$a$-C, which is
linked directly to the optical response provides information about the
disorder in the structure. We have recently shown \cite{Patsalas}, that E$_U$ is a good measure of disorder in $a$-C, as it reflects not only bond-length and bond-angle distortions (structural disorder), but also topological disorder related to the size distribution of chains
  and clusters of sp$^2$ and sp$^3$ atoms in the mixed phase
  \cite{Robertson_rev}.  Thus, it is a measure of the inhomogeneous
  disorder in the two-phase network. For example, in highly
  tetrahedral t$a$-C, the larger the sp$^2$ cluster sizes the higher
  the E$_U$ values. This is because the embedding of minority
  configurations into a host phase produces disorder, which is
  reflected in the $\pi$-$\pi^\star$ subgap transitions. We also
  showed \cite{Patsalas} that transitions between states localized on
  unpaired sp$^2$ sites (dangling bonds), or between such states and
  neighboring $\pi$ or $\sigma$ states, are negligible. This is in
  contrast to a-Si:H, where dangling bond defects are the main
  contributors to the Urbach edge.

E$_U$ for a typical composite is shown in
Fig.\ref{fig:urbach} and underlines a nonmonotonic behavior for both the
composite and pure $a$-C. The variation of E$_U$ in the case of
$n$D/$a$-C comes from the atoms in the matrix, but with higher values
at the maximum of the variation and at the tetrahedral region,
indicating the excess disorder produced on these atoms. We have
also broken E$_U$ down into atomic components by calculating
separately the absorption coefficient ($\alpha$) contributed by
the atoms close to the diamond nucleus and close to the
interface. Results shown in Fig.\ref{fig:urbach} indicate that the
variation in $\alpha$ nearly coincides with the contribution from the
atoms at the interface, while the absorption from the nucleus
approaches the absorption of crystalline diamond. The atoms at the
interface determine the E$_U$ for the whole composite. Defects at the
interface (unpaired sp$^2$ sites and dangling bonds) contribute
negligibly to the optical transitions and the Urbach edge, as in
pure $a$-C \cite{Patsalas}. On the other hand, $E_U$ for the diamond nanoinclusions takes quite high values ($\sim$ 0.6) and it is nearly constant as a function of sp$^3$ fraction. This shows the significant structural disorder at the outer shell of the nanocrystals.

\section{Summary\label{sec:concl}}

In this review, we have briefly presented some of our work on the
energetics, stability, mechanical and optoelectronic properties of
carbon-based nanostructured composites. These are materials consisting of a carbon
crystalline nanoinclusion embedded in an amorphous
carbon matrix. The inclusion ranges from a diamond crystallite to
sp$^3$ and sp$^2$ crystalline conformations with negative and positive
curvature.  The ability to control the mechanical and optoelectronic
properties of these materials by varying the density of the embedding
medium and the type and size of the crystalline inclusion opens up
enormous possibilities for potential technological applications of
carbon nanocomposites.

We found diamonds and schwarzites to be highly stable in dense
amorphous carbon matrices, while the other sp$^3$ and sp$^2$
conformations are metastable. A considerable enhancement of the
elastic moduli for $n$D/$a$-C and CNT/$a$-C was found as compared to
pure $a$-C values with increasing density and inclusion size. $a$-C
and $n$D/$a$-C of high density showed perfect elastic recovery, which
was not the case for CNT/$a$-C. Beyond the elastic regime, $n$D/$a$-C
fracture inter-grain and in a manner similar to pure $a$-C. Regarding
the optoelectronic properties of $n$D/$a$-C, the matrix was shown to
have a dominant role in forming both the EDOS and the dielectric
function, with increasing sp$^3$ fraction and density. Analysis of the
optical response and the Urbach energy, showed that much of the
disorder in the composites originates from the interface region.
\vspace{0.5cm}\\
{\bf Acknowledgments}\\
The authors wish to thank G. Kopidakis for useful discussions. % and granting access to his TBMD code.
This work was supported by the Ministry of National Education 
and Religious Affairs of Greece through the action
"$E\Pi E A K$" (program "$\Pi Y \Theta A \Gamma O P A \Sigma$".) MF acknowledges support from the 'Gender Issue Incentive Funds' of the Cluster of Excellence in Munich,
 Germany.

%\newpage
\clearpage

{\Large {\bf List of Table Captions}}

\begin{table}[h!]
 \begin{center}
 \caption{Computational details of the composites with the various inclusions presented in this review. N$_{tot}$, are the total number of atoms. N$_{cr}$, and N$_{am}$ are the number of atoms of the crystalline inclusion and the amorphous matrix, respectively. d and $\lambda$ are the initial inclusion diameter and the filling ratio. Note, that not all inclusions are spherical; for these, d gives a rough estimate of their size. Next to the CNTs the chirality indices (n,m) are given.}
\label{Tab:struc}
\end{center}
\end{table}

\begin{table}[h!]
 \begin{center}
 \caption{Elastic moduli (in GPa) for SWNT nanocomposites of different densities
$\rho$ (in gcm$^{-3}$) and nanotube diameters d (in nm).}
\label{Tab:modCNT}
\end{center}
\end{table}

\clearpage

{\Large {\bf List of Tables}}

\begin{table}[h!]
 \begin{center}
 \begin{tabular}{|c||c|c|c|c|c|}
 \hline
Inclusion &N$_{tot}$&N$_{cr}$&N$_{am}$&d(nm)&$\lambda$(\%) \\ \hline \hline
D&4096&464&3632&1.70&11\\ \hline 
BC8&2744&325&2419&1.48&12\\ \hline 
6.8$^2D$&3000&329&2671&1.79&11\\ \hline 
PCCM&2592&304&2288&1.68&12\\ \hline 
C$_{168}$&4536&636&3900&2.66&14\\ \hline 
SWNT: (6,6)&460&144&316&0.82&31\\ \hline 
SWNT: (7,10)&1486&292&1194&1.17&20\\ \hline 
SWNT: (9,9)&1362&288&1074&1.24&21\\ \hline 
SWNT: (12,12)&2097&480&1617&1.65&23\\ \hline 
SWNT: (15,15)&2828&660&2168&2.06&23\\ \hline 
SWNT: (20,20)&4910&1120&3790&2.75&23\\ \hline 
MWNT(2 walls): (10,10)@(15,15)&3409&1177&2232&2.06&34\\ \hline 
MWNT(3 walls): (5,5)@(10,10)@(15,15)&3645&1413&2232&2.06&37\\ \hline 
\hline
\end{tabular}
%\label{Tab:struc}
\end{center}
Table 1
\end{table}

\begin{table}[h!]
 \begin{center}
  \begin{tabular}{|c|c|c|c|c|c|c|}
 \hline
 d(nm)&$\rho$&$C_{11}$&$C_{12}$&$C_{44}$&Y&B(GPa)\\ \hline \hline
1.17&2.22&337&121&79&273&193 \\ \hline
1.17&2.41&418&104&108&377&209 \\ \hline
1.65&1.99&290&78&71&257&149\\ \hline
1.65&2.24&437&73&131&416&194\\ \hline
2.75&1.73&251&72&82&219&132 \\ \hline
2.75&1.84&331&61&120&311&151\\ \hline
\hline
\end{tabular}
%\label{Tab:modCNT}
\end{center}
Table 2
\end{table}

\clearpage

{\Large {\bf List of Figure Captions}}

\begin{figure}[h!]
\begin{center}
\caption{Ball and stick models of carbon nanocomposites: (a) diamond,
  (b) schwarzite $C_{168}$, and (c) a (7,10) SWNT, embedded in $a$-C
  matrices of high (for diamond) and low (for $C_{168}$ and the SWNT)
  densities. The atoms of the matrix and the embedded crystallite are indicated by white and black circles, respectively.}
\label{fig:strucs}
\end{center}
\end{figure}

\begin{figure}[h!]
\vspace*{-1cm}
\begin{center}
\caption{Formation energies for all carbon composites studied
  here. Panel (a) shows the n$a$-C, apart from the SWNTs which are
  shown in (b) for various diameters denoted in the legend. Labels are self-explaining
  according to the text. Lines are fits to the points. }
\label{fig:eform}
\end{center}
\end{figure}

\begin{figure}[h!]
\vspace*{-1cm}
\begin{center}
\caption{(a) Bulk ($B$) and Young's ($Y$) moduli, and ({\bf b}) elastic constants for
  $n$D/$a$-C as functions of the total density ($\rho$). The average
  diameter of the diamond inclusion is 1.7 nm.}
\label{fig:nDmodul}
\end{center}
\end{figure}

\begin{figure}[h!]
\vspace*{-1cm}
\begin{center}
\caption{(a) Bulk moduli for CNT/$a$-C of various diameters (shown in the legend) as functions of the density of
  the composite. The dotted line is a fit to the
  $a$-C data. (b) Comparison of $B$ for SWNT and MWNT composites with
  diameters of 2.06 nm.}
\label{fig:swntmodul}
\end{center}
\end{figure}

\begin{figure}[h!]
\vspace*{-1cm}
\begin{center}
\caption{Left panel: Stress vs strain curves for structures under
  tensile load along the \{111\} direction (see text for details).
  Right panel: Ball and stick model for fracture in the $n$D/$a$-C for
  which the stress vs. strain curves are shown on the left. Atoms are
  shaded according to the legend. Large spheres (with label (b))
  denote atoms with broken bonds. The snapshot corresponds to the
  maximum stress, at which bonds have started to break.}
\label{fig:fract}
\end{center}
\end{figure}

\begin{figure}[h!]
\vspace*{-1cm}
\begin{center}
\caption{EDOS (left) and calculated $\varepsilon_2$ (right) of $n$D/$a$-C
  with sp$^3$ fractions and densities (a) 88 \% and 3.24 g cm$^{-3}$,
  (b) 71\% and 2.91 g cm$^{-3}$, and (c) 51\% and 2.58 g cm$^{-3}$,
  respectively. The black, blue, and red lines refer to the total
  function, and the inclusion and $a$-C matrix projected
  contributions, respectively. The Fermi level is at 0 eV.}
\label{fig:opt}
\end{center}
\end{figure}

\begin{figure}[h!]
\vspace*{-1cm}
\begin{center}
\caption{(a) Variation of E$_U$ in $n$D/$a$-C as a function of the
  sp$^3$ fraction in the matrix (AM). Filled circles and squares are the
  contributions from the inclusion and the matrix, respectively. Open
  circles are values from core atoms in the inclusion. Diamond and
  triangle are values for diamond and {\it amorphous diamond},
  respectively.  Light gray line shows E$_U$ in pure $a$-C. (b)
  Breakdown of $\alpha$ (dashed-dotted red line) of a $n$D/$a$-C with
  88\% sp$^3$ fraction and 3.24 g cm$^{-3}$ in the $a$-C matrix into
  contributions from the core (black solid line) and the
  interface (black dashed line) atoms. Solid brown line indicates the
  calculated absorption of diamond.}
\label{fig:urbach}
\end{center}
\end{figure}

\clearpage

{\Large {\bf List of Figures}}

\begin{figure}[h!]
\begin{center}
\epsfig{file=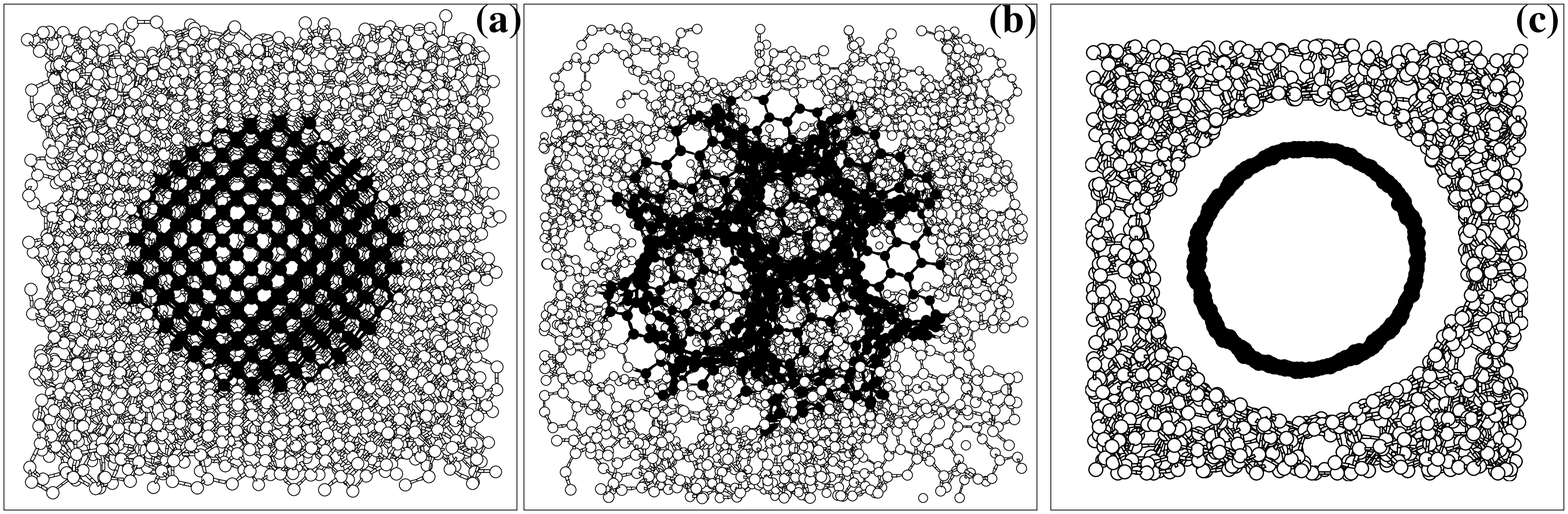,width=1\textwidth}
%\label{fig:strucs}
\end{center}
Figure 1
\end{figure}

\begin{figure}[h!]
\begin{center}
\epsfig{file=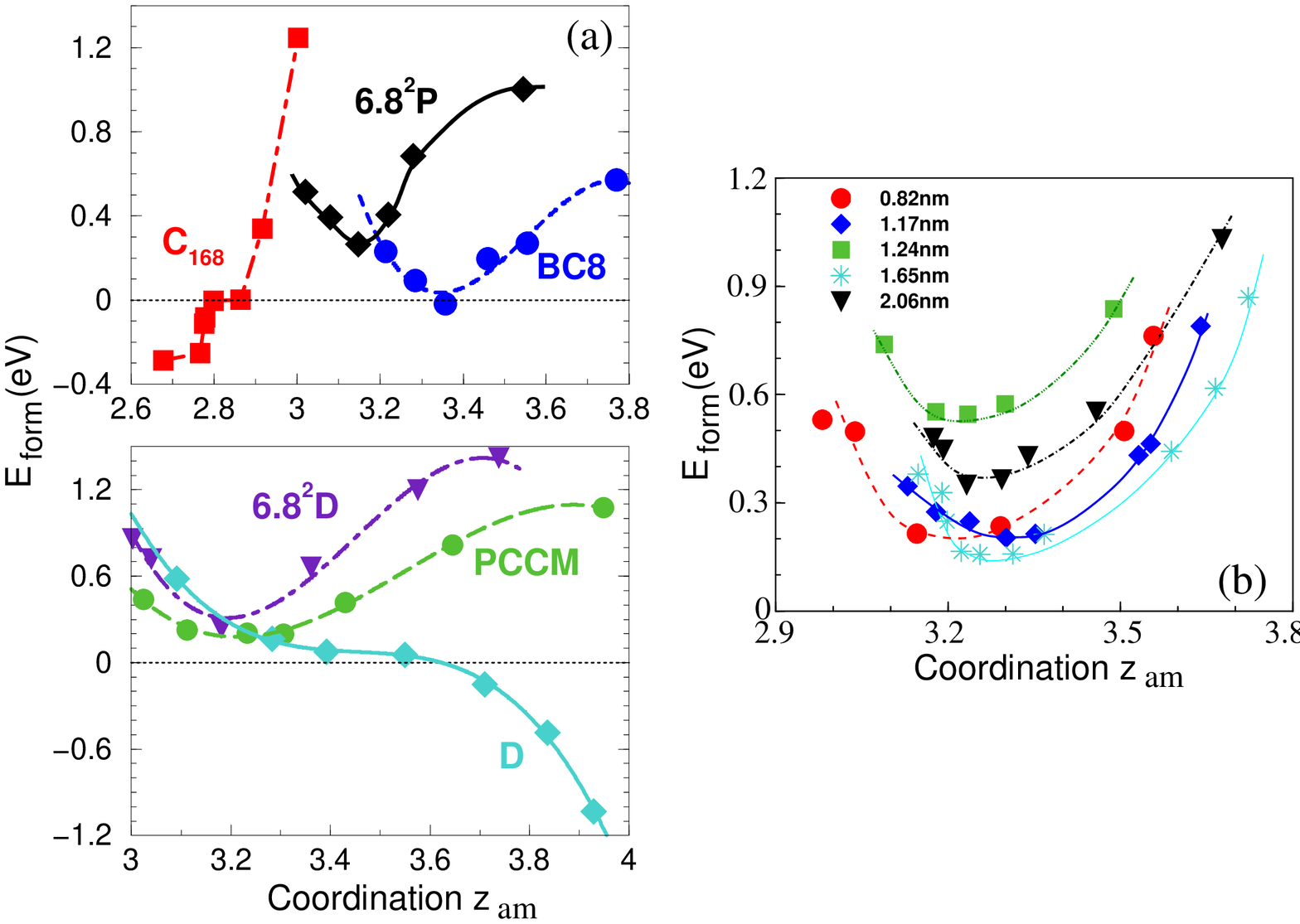,width=1.0\textwidth}
%\label{fig:eform}
\end{center}
Figure 2
\end{figure}

\begin{figure}[h!]
\begin{center}
\epsfig{file=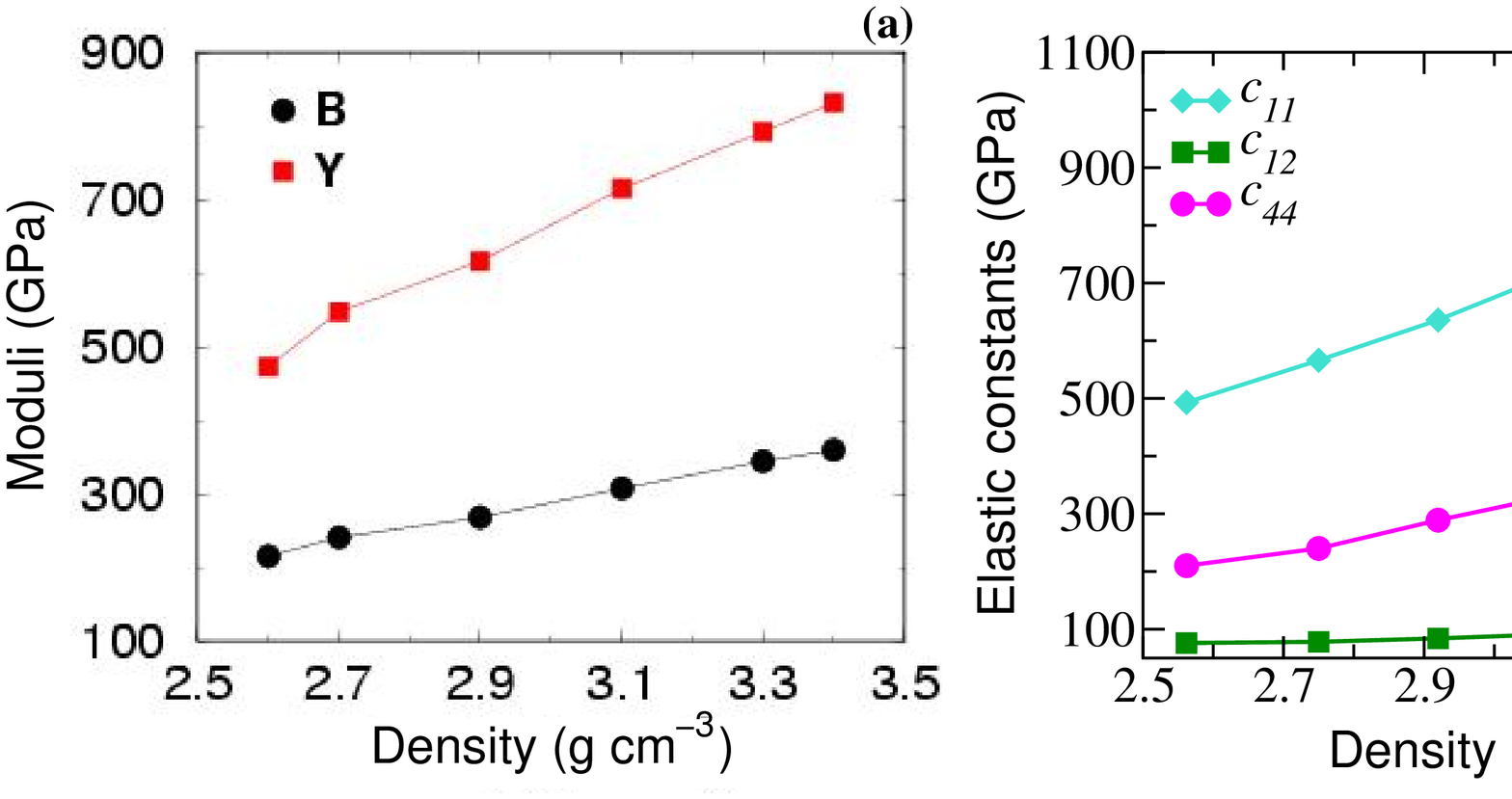,width=1\textwidth}
%\label{fig:nDmodul}
\end{center}
Figure 3
\end{figure}

\begin{figure}[h!]
\begin{center}
\epsfig{file=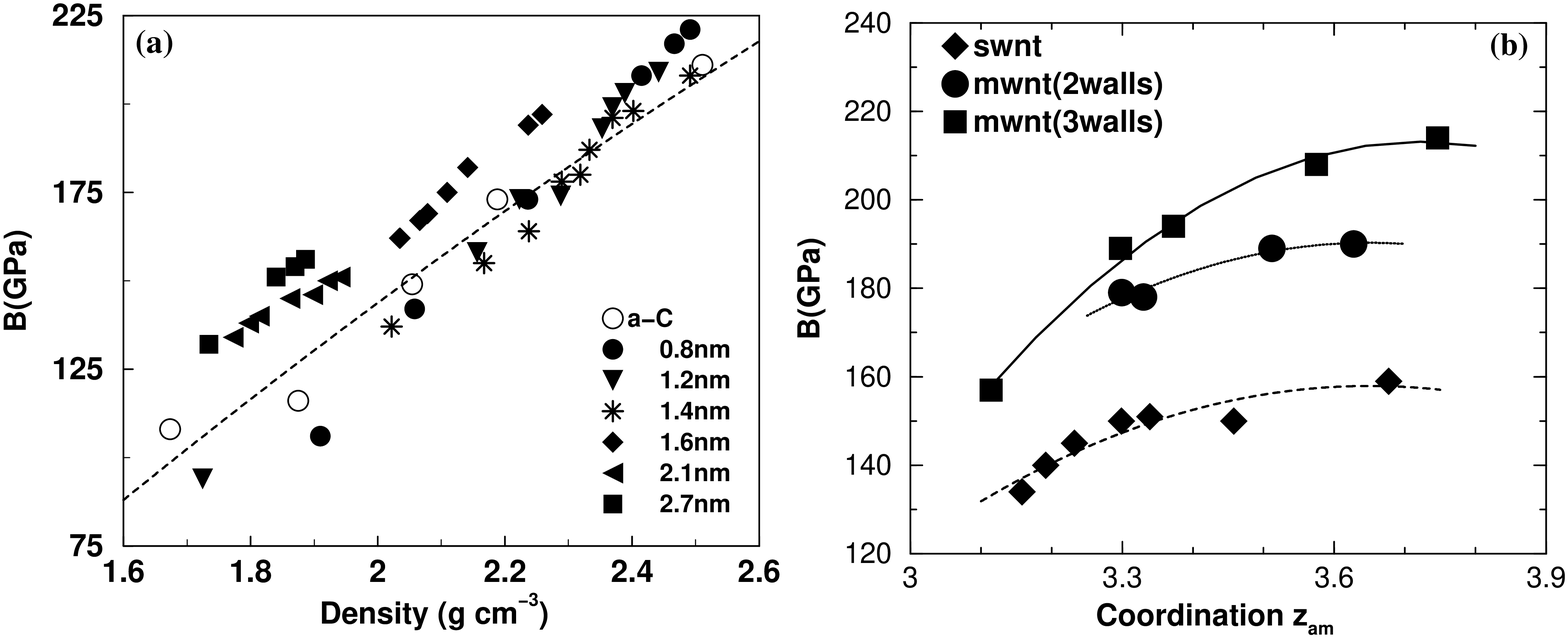,width=1\textwidth}
%\label{fig:swntmodul}
\end{center}
Figure 4
\end{figure}

\begin{figure}[h!]
\begin{center}
\epsfig{file=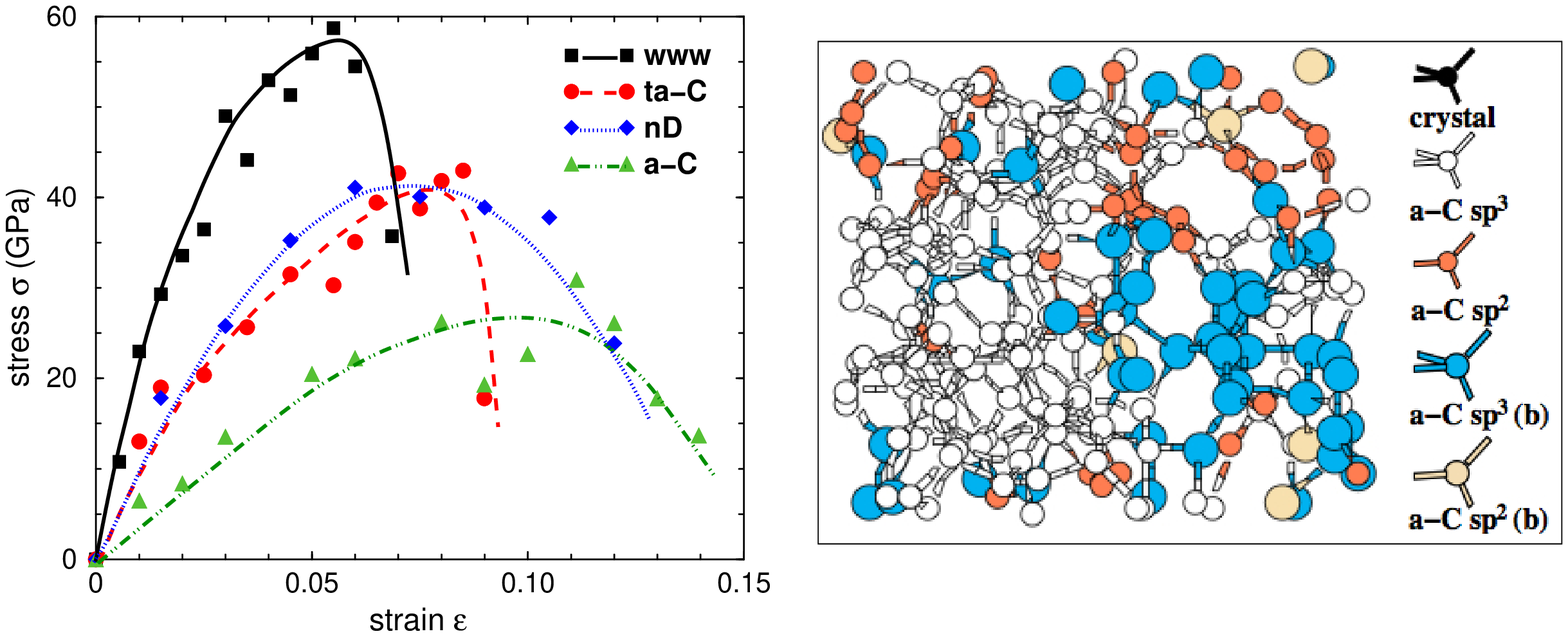,width=1\textwidth}
%\label{fig:fract}
\end{center}
Figure 5
\end{figure}

\begin{figure}[h!]
\begin{center}
\epsfig{file=edos_eps_nD_chrm_color.eps,width=0.9\textwidth}
%\label{fig:opt}
\end{center}
Figure 6
\end{figure}

\begin{figure}[h!]
\begin{center}
\epsfig{file=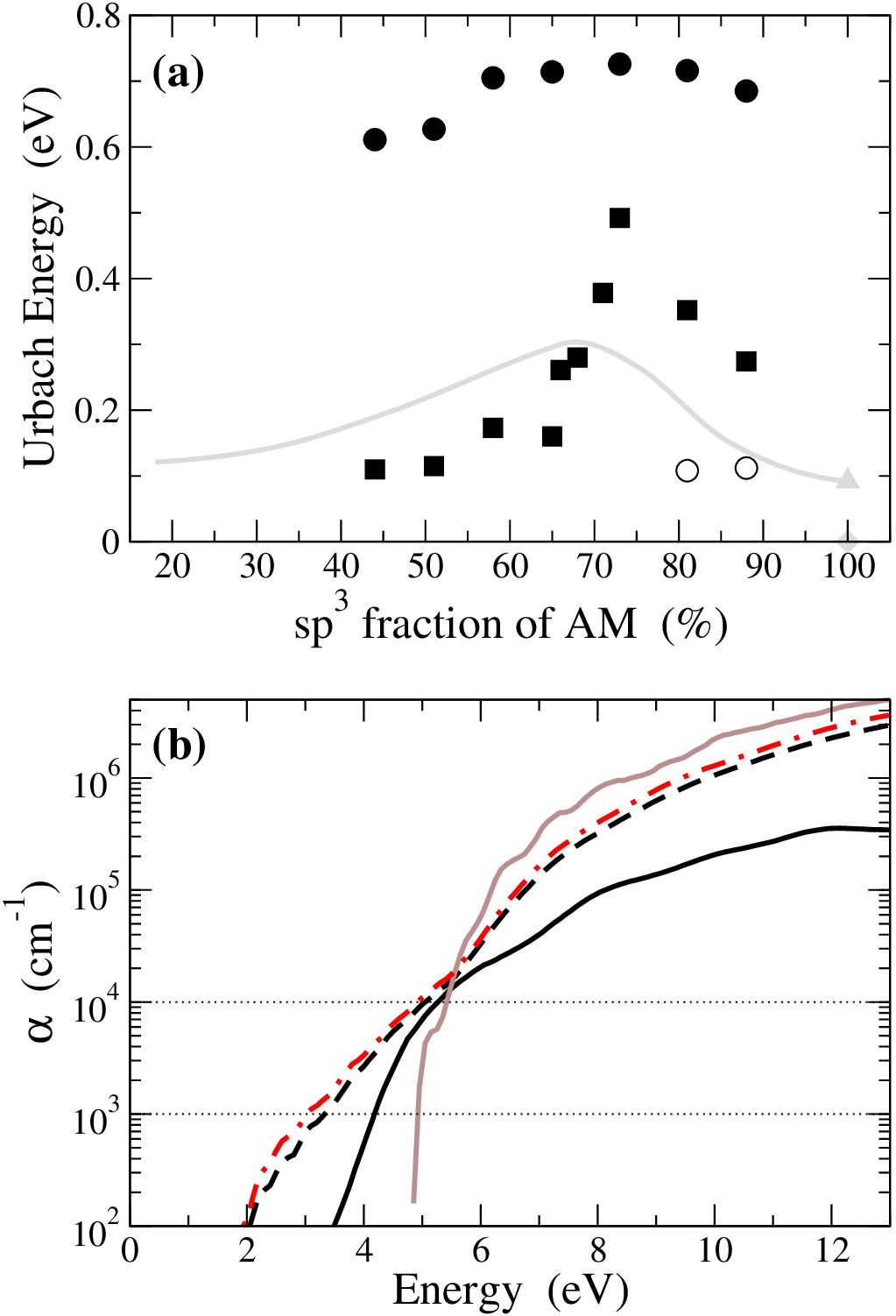,width=0.7\textwidth}
%\label{fig:urbach}
\end{center}
Figure 7
\end{figure}

\end{document}